# RESONANCE NONLINEAR REFLECTION FROM NEUTRON STAR AND ADDITIONAL RADIATION COMPONENTS OF CRAB PULSAR

V.M. Kontorovich[1,2], I.S. Spevak[3], V.K. Gavrikov[1]

[1] *Institute of Radio Astronomy NAS, Kharkiv, Ukraine*
[2] *V. N. Karazin Kharkiv National University, Kharkiv, Ukraine*
[3] *O. Ya. Usikov Institute for Radiophysics and Electronics NAS, Kharkiv, Ukraine*
E-mail: vkont@rian.kharkov.ua

Additional high-frequency components of the pulsar radiation in Crab Nebula are considered as a result of the resonance with the surface electromagnetic wave at nonlinear reflection from of the neutron star surface. This stimulated scattering consists in generation of a surface periodic relief by an incident field and diffraction on that relief the radiation of relativistic positrons, which fly from the magnetosphere to the star in the accelerating electric field of a polar gap.

## 1. INTRODUCTION

Pulsars [1] and neutron stars [2] celebrated their 50-th anniversary by international conferences in St Petersburg with a subtitle «50 years after» and in Cambridge with a subtitle «The Next Fifty Years» [1]. The materials of these conferences include reviews and modern literature references. The references to the works mentioned below are contained also in the previous article of one of coauthors [3]. We use some texts from it in the introductory part of the present work.

Neutron stars, as it is known, were predicted by Landau[2], connected with supernovae by Baade and Zwicky and discovered as pulsars by J. Bell and A. Hewish and their colleagues after a quarter of a century. Appearing as a result of the collapse at supernova explosion, the neutron star possesses the strongest magnetic field $10^{12}$ G, rapid rotation (with a period from seconds to milliseconds), and is enveloped in the *magnetosphere* of the electron-positron pairs. The magnetosphere mainly rotates corotationally with the star, but it has a bundle of open magnetic field lines over the magnetic poles; the particles are accelerated and electromagnetic radiation goes out along these lines [4]. Particles acceleration takes place in the gap under the area of open magnetic field lines where the strong accelerating electric field is caused by the magnetic field and rotation. The region of the polar cap is limited by the magnetic field lines that are tangent to light cylinder where the velocity of the corotational rotation equals to light velocity.

The neutron star itself [5] is known to correspond to the nuclear density accompanying the neutronization reaction $p + e^- \to n + \nu$. It contains layers possessing superfluidity, and possibly superconductivity. The properties of matter at such nuclear densities have not been studied sufficiently, therefore, there are a number of differing theoretical models [6].

Very little is known about the properties of the surface of neutron stars. In the case of Crab Pulsar, it, apparently, has a solid crust undergoing to starquakes. Due to the colossal force of gravity, the surface is close to mirror one, but it can contain a regular structure of elevations caused by the influence of a strong magnetic (and electric) field. In the region of the polar cap the surface can be substantially perturbed by the incident radiation. In this region, the upper layer, heated by accelerated particles and radiation, can be in the liquid state (see references and discussion [4, pp. 110; 6]). According to the references, we assume that the boundary resembles a metal with iron nuclei and collectivized degenerate electrons and it has high conductivity.

The considered mechanism of radio emission of a pulsar in the Crab nebula is based on the idea of radiation reflection from the surface of a neutron star [7]. The radiation of positrons flying to a star from the magnetosphere is reflected. This reflected radiation predo-minates in the centimeter wave range, where a shift of the interpulse (IP) occurs and high-frequency (hf) components HFC1 and HFC2, described by Moffet and Hankins, appear [8,9].

---

[1] International Conference Physics of Neutron Stars. 50 years after. St-P – 2017. Pulsar Astrophysics: The Next Fifty Years; Cambridge – 2017, IAU Symposium 337. See also jubilee review [25].

[2] It is quite natural that word *neutron* is absent in Landau's work. The neutron was discovered by Chadwick the same year and Landau could not yet know about it. For the results of Landau work the specific role of the neutron was not so significant. In his work the possibility of existence of a macroscopic atomic nucleus controlled by gravity was proved. See bright details in [26].

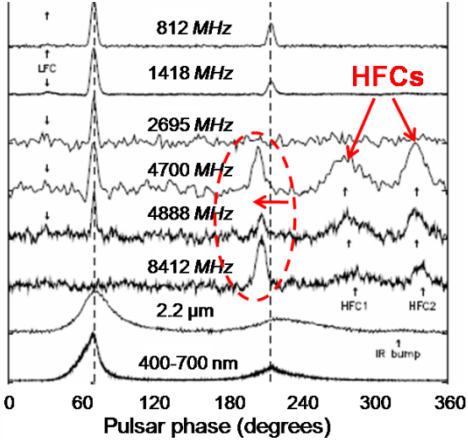

**Fig. 1**. Data of multifrequency observations of pulsar radiation in the Crab from the work of D. Moffet and T. Hankins [8] (fragment). Courtesy to the authors.

We consider the observed shift of the IP as an argument in favor of good reflecting properties of the surface. The reciprocal motion of positrons arises in the accelerating electric field of the gap and was considered earlier in the context of star surface heating.

Most researchers believe that radio emission occurs in the interior of the magnetosphere or beyond, near the "light cylinder" [1, 4, 10]. We are interested in radiation emanating from the inner gap above the polar cap, for which there are a number of arguments.

In the model [7] the displacement of the IP is explained by the mirror reflection in the inclined magnetic field, and the shift of the IP to $7°$ means the slope of the field by $3.5°$, which we will use below. The appearance of hf-components, according to [3], can be due to a non-linear reflection[3] consisting in the diffraction of the incident radiation by a periodic structure arising from the mixing of this radiation with a "material" surface wave (MSW), which is also excited by it. For definiteness, we give expressions for gravitational waves on a liquid surface.

The wave of "radiation pressure" arising at the boundary, bilinear in the amplitudes of the incident $E_0 \propto exp[i(\mathbf{kr}-\omega t)]$ and scattered $E_{\pm 1} \propto exp[i(\mathbf{k} \pm \mathbf{q})\mathbf{r} - i(\omega \pm \Omega)t]$ combinational[4] waves [14, 15],

$$p_{rad} \propto E_0 E_{-1}^* + E_0^* E_1 \propto \exp(i\mathbf{qr} - i\Omega t), \quad (1.1)$$

---

[3] Another alternative point of view of S.A. Petrova [11] is to attract induced Compton scattering of narrow beam of pulsar radiation in the depth of magnetosphere.

[4] Combinational (Raman) scattering at the surface was considered in the classical papers by Mandelshtam, Andronov and Leontovich [12, 13].

in turn, swings the surface oscillations, leading to *stimulated scattering* (SS).

For a liquid medium, solving the linearized equations of motion of the incompressible liquid

$$\rho \partial \mathbf{v} / \partial t = -grad\ p + \rho \mathbf{g},$$

taking into account the forces of electromagnetic fields, with boundary conditions at $z = \zeta$ (see [16]):

$$p^{II} - p^I - \alpha(\partial^2/\partial x^2 + \partial^2/\partial y^2)\zeta = p_{rad},$$
$$p_{rad} \equiv \Pi_{nn}^I - \Pi_{nn}^{II}, \quad \partial \zeta/\partial t = v_n,$$

where $p \equiv p' - \rho(\partial \varepsilon/\partial \rho)E^2/8\pi$, $\Pi_{nn}$ is the normal component of the Maxwell stress tensor, $p'$, $\mathbf{v}$, $\rho$ and $\alpha$ are the pressure, velocity, density of star (near the surface) and surface tension coefficient, $\mathbf{g}$ is the gravitational acceleration, we find the Fourier component of the surface oscillation $\zeta_{\mathbf{q}\Omega}$. It is expressed in terms of the Fourier component of the radiation pressure

$$\zeta_{\mathbf{q}\Omega} = |q|(p_{rad})_{\mathbf{q}\Omega}/\rho\left[\Omega_0^2(q) - \Omega^2\right], \quad (1.2)$$

where $\Omega_0(q) = (gq + \alpha q^3/\rho)^{1/2}$ is the dispersion law for the gravity-capillary wave.

The pressure wave amplitude at frequency $\Omega$ is

$$p_{rad} = {}^2 iq\zeta P\varepsilon^I |E_0^i|/8\pi, \quad (1.3)$$

where $E_0^i$ is the amplitude of incident wave field, and the dimensionless pressure $P$, given in [3, 14, 15], contains dependences on the wave vectors of electromagnetic and surface waves. We find the dispersion equation for surface waves on the irradiated surface, taking into account (1.1) and (1.3) and including attenuation due to (small) viscosity $\nu = \eta/\rho$ in it:

$$\Omega(q) = \pm\Omega_0(q) - 2iq^2\nu \mp \frac{iq^2P\varepsilon^I|E_0^i|^2}{16\pi\rho\Omega_0(q)}.$$

At the intensity of the incident field, larger than threshold, $I_0 > I_{th}$, that is

$$\varepsilon^I \frac{|E_0^i|^2}{8\pi} > \frac{4\eta\Omega_0(q)}{|\operatorname{Re} P|}, \quad (1.4)$$

growth of surface waves and stimulated scattering at them take place. The analysis is reduced to investigation of the quantity $\operatorname{Re} P$ proportional to the radiation pressure.

## 2. RESONANCE WITH SURFACE ELECTROMAGNETIC WAVE

The stimulated combinational scattering under conditions of resonance with surface electromagnetic wave (SEW) was considered in the work by Kats and Maslov

[15]. The pressure analysis was carried out for a sliding wave with $k_{\pm 1z} = 0$. Index $\pm 1$ indicates a combinational (anti-Stokes or Stokes) wave of the first order with frequency $\omega \pm \Omega$. Below we consider the same combinational waves, but we omit this index in dimensionless quantities. As follows from the general expressions for scattering fields, small denominators of the form $1/(k_{1z} - k/\sqrt{\varepsilon})$ arise in light pressure in the vicinity of the sliding scattered wave with $k_{1z} = 0$ for large value permittivity $|\varepsilon| \gg 1$.

Large $|\varepsilon|$ arises at high conductivity, and at the same time $\varepsilon = \varepsilon^{II}/\varepsilon^{I}$ is a complex quantity. It is convenient to use a complex surface impedance $\xi$, [16], where, $\xi \propto 1/\sqrt{\varepsilon}$, $|\xi| \ll 1$, instead of $\varepsilon$. We do not discuss the possible effect of magnetic permeability here. High conductivity corresponds to the concept of a boundary as a kind of high-density metal, where the iron nuclei are surrounded by free electron gas [5,6]. Then the small denominator acquires a pole view, $1/(\beta + \xi)$, where $\beta = k_{1z}/k$ (see Fig.2 and App. A).

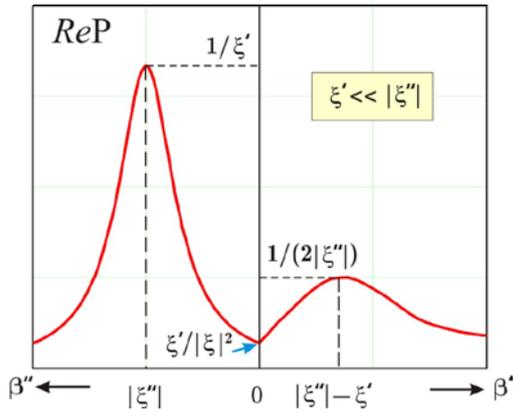

**Fig. 2.** Wood anomalies at resonance. Dependence of the real part of the light pressure on the dimensionless transverse wave number $\beta = \beta' + i\beta''$ for good conductors, ($\xi' \ll |\xi''|$). The left-hand side of the figure ($\beta' = 0$) corresponds to resonant diffraction with excitation of the SEW, the right-hand side ($\beta'' = 0$) corresponds to the near-surface diffracted wave and the maximum of total losses; at the Rayleigh point $\beta = 0$, there is a local minimum $\text{Re}\, P$ [15] in contrast to dielectrics ($\xi' \gg |\xi''|$), in which the maximum corresponds to this point [14] (see Figs 4-6 in Appendixes).

a) **Forward scattering**.

Further on, we measure all the wave numbers in units of the wave number of the incident wave $k$. Then $\beta = \sqrt{1 - (k_x + q)^2}$. Here $q$ is the wave number of the MSW, $k_x = \sin\theta$. The hf-component corresponds to the excitation of the MSW with the value of the (algebraic) wave number $-q$ and the combinational scattering at it (for details see [3]). There is no fundamental difference from real values $\varepsilon$ here. We explain the large width of the HFCI component below.

b) **Backward scattering**.

In this case, we denote the wave number of the MSW through $g$; $\beta = \sqrt{1 - (k_x + g)^2}$. The presence of a pole results in maximum $\text{Re}\, P$ at purely imaginary value $\beta \equiv i\beta''$ corresponding to the resonance with SEW – the eigen wave of the surface. This purely inhomogeneous wave does not radiate into space, but we will see below that it can contribute to scattering at higher frequencies. The value of pressure at maximum $\beta'' = |\xi''|$ can be very large: $\text{Re}\, P_{max}^{res} \approx 1/\xi'$.

The second maximum $\text{Re}\, P$ arises for real $\beta \equiv \beta' = -(\xi'' + \xi')$ and corresponds to the propagating near-surface wave (see App. B): $\text{Re}\, P_{extr}^W \approx 1/\xi''$.

Finally, the special case is the Rayleigh point for $\beta = 0$, considered in [3]: $\text{Re}\, P^R \approx \sqrt{|\varepsilon|} \gg 1$. In terms of impedance $\text{Re}\, P^R \approx \xi'/|\xi|^2$.

All three characteristic features of $\text{Re}\, P$ are closely related to the so-called Wood anomalies[5] and are basically generated by the diffraction peculiarities near Rayleigh angle of incidence for a certain diffraction order.

## 3. EXPLANATION OF THE HF-COMPONENTS WIDTH

The width of the hf-components reaches $30°$, that is substantially greater than the width of the IP. This model presents a simple physical explanation for the hf-components broadening.

Suppose that SS is realized at certain frequency $\omega_1$ from the continuous spectrum of positron emission flying to the star. This means the excitation and buildup of the MSW with a certain value of the wave vector $q_1$ (for forward scattering, cf. Fig. 3 in [3]), or $g_1$ (for backward scattering, see Fig. 3 below). For a higher frequency $\omega_2$ of positron emission, a near-surface propagating electromagnetic wave arises for combinational scattering on the same MSW $g_1$ if frequency difference $\omega_2 - \omega_1$ exceeds the (small) surface impedance (Fig. 3).

Thus, there is a contribution to the component HFC2 from the wide region of the positron emission spectrum, resulting in a large width of the component. Analogously, such pulse broadening occurs for HFC1 component due to MSW $q_1$ at forward scattering. If we assume that SS at frequency $\omega_1$ occurs at the surface resonance that corresponds to non-radiated non-uniform electromag-netic wave with a purely imaginary transverse wave number, then reflected propagating waves appear at higher frequencies $\omega_2$. The set of different frequency combinations explains the large width of the hf-component. The steeply falling energy spectrum in Crab Pulsar is an important argument in favor of such scenario.

---
[5] Wood anomalies [17] received first physical explanation in the work by Rayleigh [18], who connected them with a sliding diffraction orders. See also the monograph [19].

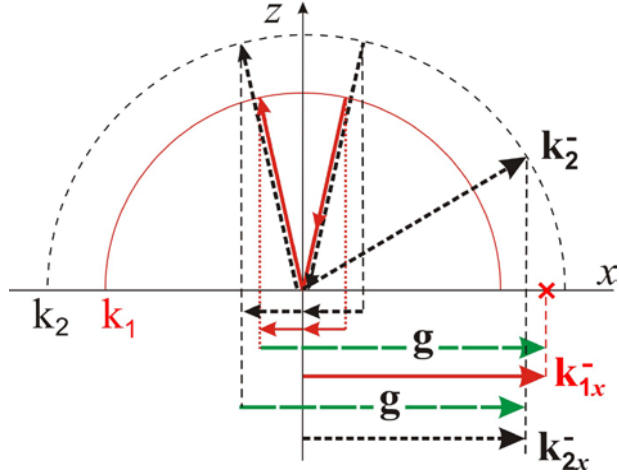

**Fig.3.** A diagram explaining the contribution to the nonlinear reflection of combinational fields at higher frequency $\omega_2$ from material surface waves ($\mathbf{g}$) excited via resonant stimulated scattering by the incident radiation at lower frequency $\omega_1$. For continuous spectrum of the radiation incident on the star this explains the large width of the hf-components. The details are in the text[6].

## 4. CONCLUSION

This work is based on the idea that the radiation of the return positrons is reflected from the surface of a neutron star, which was introduced by one of the authors and S.V. Trofimenko [7, 20-22]. The reflection in a magnetic field inclined to the surface star manifests itself in the IP shift and the appearance due to diffraction[7] of additional hf-components discovered and investigated in [8-10] by Moffett, Hankins, Eilek and Jones. The possible effect of the combinational waves resonance with SEW sliding along the surface is discussed in this paper. The possibility of forming wide hf-components due to combinational scattering of a wide spectrum of positrons radiation incident on the star surface is shown. This allows us (albeit ambiguously) to explain the observed drift of the hf-component and re-turn[8], at least partly, to coupling each of the components to its pole [3]. In particular, backward scattering in the North Pole on a periodic structure excited at resonance with SEW gives such an opportunity, since it imitates "drift" towards the North Pole.

---

[6] Let us note that the wave corresponding to $\omega_1$ is not emitted, it has a purely imaginary transverse wave number. The wave corresponding to $\omega_2 > \omega_1(1+\xi''^2/2)$ has a real transverse wave number and contributes to the reflection at frequency $\omega_2$.

[7] Here, as in [3], diffraction is considered on the periodic structure created by the incident radiation (SS) itself. Another possibility − diffraction on a periodic structure caused by the action of constant fields (electric or magnetic) requires a separate consideration.

Indeed, let the MSW with a wave number $g_1$ (Fig. 3) be excited by a wave $k_1$ at resonance with SEW. Then $g_1 = k_1\left(1+\sin\theta_N + \xi''^2/2\right)$. In scattering the wave $k_2 > k_1$ on this structure, a Stokes wave with a tangential component of the wave vector $k_{2x}^- = g_1 - k_2 \sin\theta_N$ arises. At $k_{2x}^- \leq k_2$ these waves will be propagating, and the equality $k_{2x}^- = k_2$ corresponds to $\pi/2$ angle relative to normal. At higher frequencies, this angle $\varphi_2$ is

$$\varphi_2 = \arcsin\left(k_{2x}^-/k_2\right)$$

or

$$\sin\varphi_2 = k_1\left(1+\sin\theta_N + \xi''^2/2\right)/k_2,$$

from which it can be seen that the angle decreases with increasing frequency $k_2$. This corresponds to the observed "drift" of the component toward the N-pole [9].

At the South Pole (cf. the Fig.3 in [3]) a similar "drift" may occur due to scattering of lower frequencies and the angle $\varphi_2^S$ will also increase in accordance with observations, but this will require the excitation of a wide range of material surface waves.

## APPENDIX A

Consider the SEW (surface electromagnetic wave) in terms of the surface impedance. From the Maxwell equation for the H-wave $\operatorname{rot}_x \mathbf{H} = -i\varepsilon^I(\omega/c)\mathbf{E}_x$, and the Leontovich boundary condition

$$\mathbf{E}_x = \sqrt{\mu/\varepsilon^{II}}\,[\mathbf{H},\mathbf{n}]_x,$$

$\mathbf{n}$ is the normal to the surface, follows:

$$\partial H_y/\partial z = -ik\xi H_y.$$

where $k$ is the wave number of electromagnetic wave in the first medium and $\xi$ is a relative impedance.

SEW corresponds $H_y \propto \exp(-\kappa z)$ with $\kappa > 0$ (the axis $z$ is directed into the first medium) whence the dispersion relation for SEW takes the form

$$\kappa = ik\xi,$$

which requires a purely imaginary impedance $\xi = i\xi''$, $\xi'' < 0$, corresponding to the negative $\varepsilon^{II} < 0$ (if $\mu$ and $\varepsilon^I$ are real and positive) [16]. By introducing a dimensionless transverse component of the wave vector, $\beta$, according to $k_z \equiv \beta k$ or $\kappa \equiv -i\beta k$ with $\text{Im}\,\beta \geq 0$, we rewrite the condition in the form

$$\beta + \xi = 0.$$

---

[8] We note that in [22], just as in S.A. Petrova's works, the origin of two HF components was considered as a result of birefringence in a magnetospheric plasma.

We emphasize that the SEW as the eigen wave exists for real and negative value of $\varepsilon^{II}$ only (with the remark made above). In the general case of the complex $\varepsilon^{II}$, small denominators of the form $(\beta + \xi)$ arise in the diffracted fields in the region of Wood anomalies. Respectively, resonance combinational field acquire a form

$$H^{\pm 1} \propto 1/(\beta + \xi).$$

Then the resonant part of the dimensionless radiation pressure, bilinear in the incident and combinational field, is

$$P \propto 1/(\beta + \xi),$$

and the real part $\text{Re}\,P$, which determines the instability increment, is equal to

$$\text{Re}\,P = \frac{\beta' + \xi'}{|\beta + \xi|^2} = \frac{\beta'}{|\beta + \xi|^2} + \frac{\xi'}{|\beta + \xi|^2}.$$

The above expression has a simple physical interpretation. The first term is proportional to the normal component of the energy flux density of the resonance spectrum

$$S_{out}^{\pm} \propto \beta' |H^{\pm}|^2 \propto \beta'/|\beta + \xi|^2,$$

carried away from the surface. The second term corresponds to the energy flux density directed to the surface,

$$S_{in}^{\pm} \propto \text{Re}[\mathbf{E}^*\mathbf{H}]_n^{\pm} \propto \xi'/|\beta + \xi|^2,$$

and this flux is completely absorbed by the medium. Thus, $\text{Re}\,P$ is proportional to the sum of radiative and dissipative losses. Accordingly, the maxima of $\text{Re}\,P$ correspond to the maxima of the total losses of the resonance spectrum. The corresponding extremal and singular points were given at the end of Section 3.

Note for certainty, the radiation (light) pressure used above is found from the solution of the electrodynamic problem using Leontovich boundary condition at the surface, perturbed by a material wave [14-15, 3].

## APPENDIX B

We illustrate the influence of the sliding waves and Wood anomalies by observing the reflection from diffraction grating (according to the data of [23]).

The mechanism of nonlinear reflection of radiation from the surface of a neutron star, described above, has much in common with the mechanism of the resonance diffraction of electromagnetic radiation on the periodic surface of the conducting medium in the vicinity of Wood anomalies. As in the case considered above, the interaction between the diffraction components of the beam leads to an increase in the intensity of Stokes (or anti-Stokes) components, resulting in a bright near-surface wave and, correspondingly, decreasing of the intensity of the specularly reflected radiation (Fig. 4).

We present some results of the laboratory studies of the resonance diffraction of radiation on a corrugated metal surface [23]. The radiation source was HCN laser (wavelength $\lambda = 366.7\,mkm$), with beam radius at the laser output $6.7\,mm$, and beam divergence $\approx 0.9\,\text{deg}$ in $1/e$ intensity level.

The experiments were carried out with brass samples ($Cu\ 60\%$), periodic structures (gratings) were prepared on their surfaces with grooves of different depths $h$: 16, 24 and $40\,mkm$. The periods $d$ of all gratings are the same: $d = 254\,mkm.$

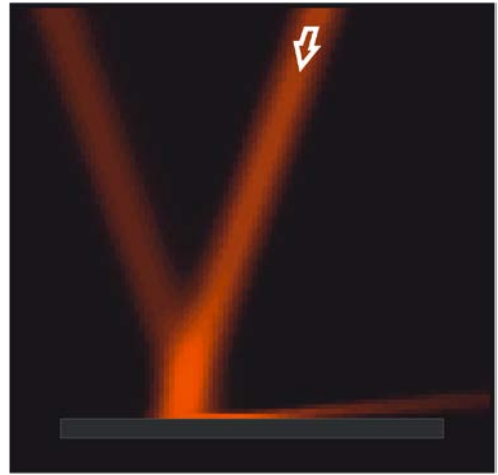

**Fig.4.** Sliding near-surface wave against a background of incident and reflected waves (numerical simulation). The bright diffracted wave corresponds to a weakened reflected one [23]. Thanks to M. Tymchenko

The diffraction anomalies were studied in the vicinity of the incidence angle $\theta$ corresponding to the Rayleigh angle for the $-1$-th diffraction order: $\theta \approx \theta_R^{(-1)} \equiv \arcsin[(\lambda/d) - 1]$. This geometry is preferable, since in this case there are only two propagating waves – specularly reflected and minus-first diffraction component. The remaining diffraction orders are inhomogeneous and have not been recorded in the experiment. The power of radiation was measured depending on the angle of incidence. At angles of incidence lower than Rayleigh one ($\theta < \theta_R^{(-1)}$), when Stokes wave $H_{-1}$ was inhomogeneous, power of the specularly reflected radiation was measured, and for angles $\theta > \theta_R^{(-1)}$ the dependence of Stokes component power on the sliding angle $\psi$ (between the grating plane and the recorded beam) was measured.

As follows from the theoretical consideration of the problem, the change in the power of the specularly reflected radiation has a quasi-resonance character. Near Rayleigh angle, the specularly reflected radiation is substantially suppressed, and the suppression increases with the deepening of the grating grooves. Respectively, the intensity of Stokes component increases. It is also seen that minimum of specular reflection at the incidence angle $\theta_{min}$ corresponds to Stokes component maximum at sliding angle $\psi_{max}$ (in case of diffraction in the –1-st order, these angles are related by $\cos\psi_{max} + \sin\theta_{min} = \lambda/d$ ). It should be noted that the shape of the curve depends on the depth of the grating grooves. For shallow gratings ( $h << d$ ) the angular dependence of reflectivity is almost symmetric and close to Lorentz curve. At the increase of the groove depth the symmetry of the wings disappears and the shape of the curve approaches a form characteristic for Fano resonance [24]. This is stipulated by the presence of two channels for signal formation: nonresonance (Fresnel) and resonance ones, caused by interaction of the diffracted order with the grating. The manifestation of this interaction is the above-mentioned redistribution of energy between the specularly reflected and Stokes components: the incident wave scattering at the grating generates Stokes component; in turn, scattering of the Stokes wave itself produces another component which propagates in the same direction as the specular reflected wave. Interference of this "additional" component with the specular reflected wave results in redistribution of energy between the diffraction components (which is illustrated by numerical simulation), and a change in the shape of the Fano resonance curve.

## APPENDIX C

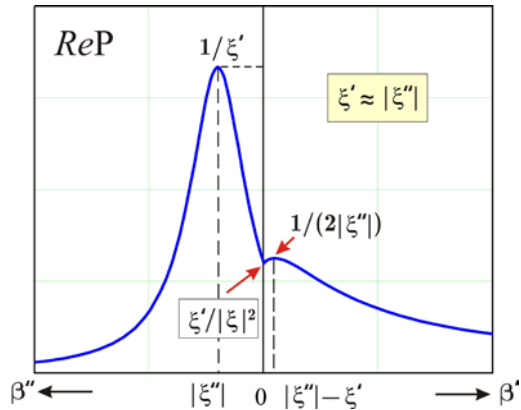

**Fig.5.** Wood anomalies in light pressure for typical metal $\varepsilon \approx 4\pi i\sigma/\omega -$ (many variants).

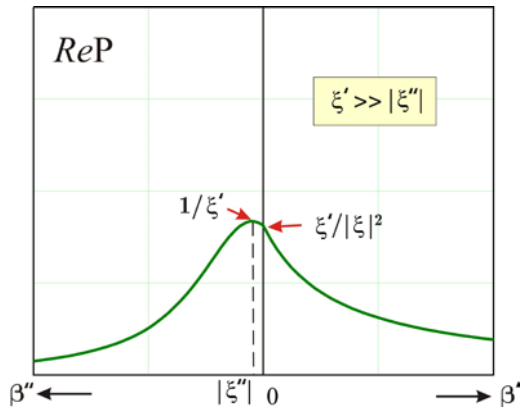

**Fig.6.** Wood anomalies in light pressure for poor conductors and dielectrics

## ACKNOWLEDGMENTS

One of the authors (V.K.) is thankful to D.M. Vavriv for help and support. We also are grateful to A.V. Kats and S.V. Trofymenko who carefully read the manuscript and made useful comments.

P.S. The contents of this e-print were partly published in articles [27, 28].